# Angular distributions of secondary electrons in fast particle-atom scattering

M. Ya. Amusia[1, 2], L. V. Chernysheva[2], and E. Z. Liverts[1]

[1]Racah Institute of Physics, the Hebrew University, Jerusalem 91904, Israel
[2]Ioffe Physical-Technical Institute, St.-Petersburg 194021, Russia

**Abstract**

We present the angular distribution of electrons knocked out from an atom in a fast charge particle collision at small momentum transfer. It is determined not only by dipole but also by quadrupole transitions, the contribution of which can be considerably enhanced as compared to the case of photoionization.

There the non-dipole parameters are suppressed as compared to the dipole ones by the parameter $\omega R/c \ll 1$, where $\omega$ is the photon energy, $R$ is the ionized shell radius and $c$ is the speed of light. This suppression in fast electron-atom collisions can be considerably reduced: the corresponding expansion parameter $\omega R/v \ll 1$ is much bigger than in photoionization, since the speed of the incoming electron $v$ is much smaller than $c$. In formation of the angular distribution it is decisively important that the ionizing field in collision process is longitudinal, while in photoionization – it is transversal.

We illustrate the general formulas by concrete results for outer s- , p- , and some *nd*-subshells of multi-electron noble gas atoms Ar, Kr and Xe, at several transferred momentum values: $q$=0.0, 0.1, 1.1, 2.1. Even for very small transferred momentum $q$, i.e. in the so-called optical limit, the deviations from the photoionization case are prominent.

1. Introduction

About ten - fifteen years ago, a lot of attention has been given to investigation of the so-called non-dipole parameters of the photoelectrons angular distribution (see [1-3] and references therein). It was understood that this is in fact the only way to reveal the contribution of quadrupole continuous spectrum matrix elements of atomic electrons that in the absolute cross photoionization cross-section are unobservable in the shadow of much bigger dipole contribution. To study non-dipole parameters high intensity sources of continuous spectrum electromagnetic radiation were used [4-7].

By the order of magnitude the ratio quadrupole-to-dipole matrix elements in photoionization characterize the parameter $\omega R/c$, where $\omega$ is the photon energy, $R$ is the ionized shell radius and $c$ is the speed of light. For photon energies up to several keV that includes ionization potential of the inner 1s subshell even for medium atoms, one has $\omega R/c \ll 1$. In the absolute cross-sections, dipole and quadrupole terms do not interfere, so that the ratio of quadrupole to dipole contributions in the absolute cross section is given by the second power of the parameter $\omega R/c \ll 1$ and some of these terms are canceling each other. As to the angular distribution, it includes the dipole – quadrupole interference terms in the first power of $\omega R/c \ll 1$ and therefore the relative role of quadrupole terms are much bigger.

Quite long ago fast charged particle inelastic scattering process was considered as a "synchrotron for poor" [8]. This notion reflects the fact that fast charge particle inelastic scattering is similar to photoionization, since it is mainly determined by the dipole contribution. But contrary to the photoionization case the ratio "quadrupole-to-dipole" contributions can be much bigger, since instead of $\omega R/c \ll 1$ they are determined by the parameter $\omega R/\upsilon$, where $\upsilon$ is the speed of the projectile. Since $1 \ll \upsilon \ll c$, the quadrupole term in inelastic scattering is relatively much bigger[1]. The transferred in collision momentum $q$ is not bound to the transferred energy $\omega$ by a relation similar to $\omega = aq$, with $a$ being a constant. Therefore, the collision experiment gives an extra degree of freedom to control the atomic reaction to the transferred energy and linear momentum. This stimulates the current research. Its aim is to derive formulas for the angular anisotropy parameters of electrons emitted off an atom in its inelastic scattering with a fast charged projectile. We perform also calculations of these parameters as functions of $\omega$ and $q$. Note that the information from photoionization studies does not inform at all the $q$-dependences of dipole and quadrupole matrix elements and the monopole matrix elements.

In this paper, we suggest to investigate the cross-section of inelastic scattering upon atom and to study the angular distribution of the emitted electrons relative to the momentum $q$ transferred to the atom from the projectile. As it is well known, fast charged particle inelastic scattering cross-section is proportional to the so-called generalized oscillator strength (GOS) density. Therefore, we will concentrate in this paper on the GOS density angular distribution as a function of the direction of the atomic electron relative to the vector $\vec{q}$. We investigate the differential cross-section of inelastic scattering upon atom as a function of the angle $\theta$ between the momentum of the emitted in collision process electron and the direction of $\vec{q}$. As it is known, the fast charged particle inelastic scattering cross section is proportional to the so-called GOS-density. Thus, in this paper we study the GOS density angular distribution as a function of $\theta$.

Deep similarity between photoionization and fast electron scattering brought to a belief that not only the total cross-section, but also angular anisotropy parameters are either the same of similar. As it is shown below, this is incorrect even in the limit $q \to 0$.

In our calculations we will not limit ourselves to the one electron Hartree-Fock approximation, but include multi-electron correlations in the frame of the random phase approximation with exchange (RPAE) that was successfully applied to studies of photoionization and fast electron scattering [9, 10].

## 2. Main formulae

The cross-section of the fast electron inelastic scattering upon an atom with ionization of an electron of $nl$ subshell can be presented as [11, 12]

$$\frac{d^2\sigma_{nl}}{d\omega do} = \frac{2\sqrt{(E-\omega)}}{\sqrt{E}\omega q^2}\frac{dF_{nl}(q,\omega)}{d\omega}. \tag{1}$$

Here $dF_{nl}(q,\omega)/d\omega$ is the GOS density, differential in the ionized electron energy $\varepsilon = \omega - I_{nl}$, $I_{nl}$ is the $nl$ subshell ionization potential.

In one-electron approximation the GOS density differential both in the emission angle and energy of the ionized electron with linear momentum $\vec{k}$ from a subshell with principal quantum number $n$ and angular momentum $l$ is given by the following formula:

$$\frac{df_{nl}(q,\omega)}{d\Omega} = \frac{1}{2l+1}\frac{2\omega}{q^2}\sum_m \left|\left\langle nlms\left|\exp(i\vec{q}\vec{r})\right|\varepsilon\vec{k}s\right\rangle\right|^2. \tag{2}$$

---
[1] Atomic system of units is used in this paper: electron charge $e$, its mass $m$ and Plank constant $\hbar$ being equal to 1, $e = m = \hbar = 1$

Here $\vec{q} = \vec{p} - \vec{p}'$, with $\vec{p}$ and $\vec{p}'$ being the linear moments of the fast incoming and outgoing electrons determined by the initial $E$ and final $E'$ energies as $p = \sqrt{2E}$ and $p' = \sqrt{2E'}$, $\Omega$ is the solid angle of the emitted electron, $m$ is the angular momentum projection, $s$ is the electron spin. Note that $\omega = E - E'$, and $\varepsilon = \omega - I_{nl}$ is the outgoing electron energy.

The values of $\omega$ are limited by the relation $0 \leq \omega \leq pq$, contrary to the proportionality $\omega = cq$ for the case of photoeffect. In order to consider the projectile as fast, its speed must be much higher than the speed of electrons in the ionized subshell, i.e. $\sqrt{2E} \gg R^{-1}$. We consider the transferred to the atom momentum $q$ as small, if $qR \leq 1$.

Expanding $\exp(i\vec{q}\vec{r})$ into a sum of products of radial and angular parts and performing analytic integration over the angular variables, one obtains for GOS in one-electron Hartree-Fock approximation:

$$g_{nl,kl',L}(q) \equiv \int_0^\infty R_{nl}(r) j_L(qr) R_{kl'}(r) r^2 dr, \qquad (3)$$

where $j_L(qr)$ are the spherical Bessel functions and $R_{nl(kl')}(r)$ are the radial parts of the HF electron wave functions in the initial (final) states.

We suggest measuring the angular distribution of the emitted electrons relative to $\vec{q}$. It means that the z-axis coincides with the direction of $\vec{q}$ and hence one has to put $\theta_{\vec{q}} = \varphi_{\vec{q}} = 0$ in Eq. (2). Since we have in mind ionization of a particular $nl$ subshell, for simplicity of notation and due to energy conservation in the fast electron inelastic scattering process leading to $k = \sqrt{2(\omega - I_{nl})}$, let us introduce the following abbreviations $g_{nl,kl',L}(q) \equiv g_{kl'L}(q)$.

The GOS formulas can be generalized in order to include inter-electron correlations in the frame of RPAE. We achieve this substituting $g_{kl'L}(q)$ by modulus $\tilde{G}_{kl'L}(q)$ and the scattering phases $\delta_{l'}$ by $\bar{\delta}_{l'} = \delta_{l'} + \Delta_{l'}$, where the expressions $G_{kl'L}(q) \equiv \tilde{G}_{kl'L}(q) \exp(i\Delta_{l'})$ are solutions of the RPAE set of equations [13]:

$$\langle \varepsilon l' | G^L(\omega, q) | nl \rangle = \langle \varepsilon l' | j_L(qr) | nl \rangle +$$
$$+ \left( \sum_{\varepsilon''l'' \leq F, \varepsilon'''l''' > F} - \sum_{\varepsilon''l'' < F, \varepsilon'''l''' \leq F} \right) \frac{\langle \varepsilon'''l''' | G^L(\omega, q) | \varepsilon''l'' \rangle \langle \varepsilon''l'', \varepsilon l' | U | \varepsilon'''l''', nl \rangle_L}{\omega - \varepsilon_{\varepsilon'''l'''} + \varepsilon_{\varepsilon''l''} + i\eta(1 - 2n_{\varepsilon''l''})} \qquad (4)$$

Here $\leq F (> F)$ denotes summation over occupied (vacant) atomic levels in the target atom. Summation over vacant levels includes integration over continuous spectrum, $n_{\varepsilon l}$ is the Fermi step function that is equal to 1 for $nl \leq F$ and 0 for $nl > F$; the Coulomb inter-electron interaction matrix element is defined as $\langle \varepsilon''l'', \varepsilon l' | U | \varepsilon'''l''', nl \rangle_L = \langle \varepsilon''l'', \varepsilon l' | r_<^L / r_>^{L+1} | \varepsilon'''l''', nl \rangle - \langle \varepsilon''l'', \varepsilon l' | r_<^L / r_>^{L+1} | nl, \varepsilon'''l''' \rangle$. In the latter formula notation of smaller (bigger) radiuses of $r_<(r_>)$ of interacting electron coordinates comes from the well-known expansion of the Coulomb inter-electron interaction. The necessary details about solving (4) one can find in [14].

For differential in the outgoing electron angle GOS density of $nl$ subshell $dF_{nl}(q,\omega)/d\Omega$ the following relation are valid in RPAE

$$\frac{dF_{nl}(q,\omega)}{d\Omega} = \sum_{L'L''} \frac{dF_{nl}^{L',L''}(q,\omega)}{d\Omega} = \frac{\omega\pi}{q^2} \sum_{L'L''} (2L'+1)(2L''+1)i^{L'-L''} \times$$
$$\sum_{l'=|L'-l|}^{L'+l} \sum_{l''=|L''-l|}^{L''+l} \tilde{G}_{kl'L'}(q)\tilde{G}_{kl''L''}(q) i^{l''-l'}(2l'+1)(2l''+1)e^{i(\bar{\delta}_{l'}-\bar{\delta}_{l''})} \begin{pmatrix} L' & l & l' \\ 0 & 0 & 0 \end{pmatrix} \begin{pmatrix} l'' & l & L'' \\ 0 & 0 & 0 \end{pmatrix}$$
$$\sum_{L=|l'-l''|}^{l'+l''} P_L(\cos\theta)(-1)^{L+l}(2L+1) \begin{pmatrix} l' & L & l'' \\ 0 & 0 & 0 \end{pmatrix} \begin{pmatrix} L & L' & L'' \\ 0 & 0 & 0 \end{pmatrix} \begin{Bmatrix} L & L' & L'' \\ l & l'' & l' \end{Bmatrix}. \quad (5)$$

This expression was obtained by generalizing (2) to include RPAE corrections and performing required analytical integrations and summations over projection of the electrons angular moments *m* with the help of Mathematica programs [15, 16].

The partial value of GOS $F_{nl}(q,\omega)$ in RPAE is obtained from (5) by integrating over $d\Omega$, leading to the following expressions:

$$F_{nl}(q,\omega) = \sum_{L'} F_{nl}^{L'}(q,\omega) = \frac{4\omega\pi^2}{q^2} \sum_{L'} (2L'+1) \sum_{l'=|L'-l|}^{L'+l} [\tilde{G}_{kl'L'}(q)]^2 (2l'+1) \begin{pmatrix} L' & l & l' \\ 0 & 0 & 0 \end{pmatrix}^2 . \quad (6)$$

Note that at small *q* the dipole contribution in GOSes $F_{nl}(q,\omega)$ dominates and is simply proportional to the photoionization cross-section $\sigma_{nl}(\omega)$ [10]. To compare the results obtained with known formulas for the photoionization with lowest order non-dipole corrections taken into account, let us consider so small *q* that it is enough to take into account terms with $L',L'' \leq 2$. In this case, GOS angular distribution (5) can be presented similar to the photoionization case as

$$\frac{dF_{nl}(q,\omega)}{d\Omega} = \frac{F_{nl}(q,\omega)}{4\pi} \left\{ 1 - \frac{\beta_{nl}^{(in)}(\omega,q)}{2} P_2(\cos\theta) + q\left[\gamma_{nl}^{(in)}(\omega,q)P_1(\cos\theta) + \eta_{nl}^{(in)}(\omega,q)P_3(\cos\theta) + \varsigma_{nl}^{(in)}(\omega,q)P_4(\cos\theta)\right] \right\}. \quad (7)$$

The obvious difference is the *q* dependence of the coefficients and an extra term $\varsigma_{nl}^{(in)}(\omega,q)P_4(\cos\theta)$. Even in this case, expressions for $\beta_{nl}^{(in)}(\omega,q)$, $\gamma_{nl}^{(in)}(\omega,q)$, $\eta_{nl}^{(in)}(\omega,q)$, and $\varsigma_{nl}^{(in)}(\omega,q)$ via $g_{kl'L'}(q)$ are too complex as compared to relations for $\beta_{nl}(\omega)$, $\gamma_{nl}(\omega)$, and $\eta_{nl}(\omega)$ in photoionization. Therefore, it is more convenient to present the results for *s*, *p*, and *d* subshells separately. We demonstrate that while $F_{nl}(q,\omega) \sim \sigma(\omega)$, similar relations are not valid for the anisotropy parameters.

Let we start with s-subshells, where as it follows from (5) the following relation gives differential GOSes in the above-mentioned $L',L'' \leq 2$ approximation

$$\frac{dF_{n0}(q,\omega)}{d\Omega} = \sum_{L',L''=0}^{2} \frac{dF_{n0}^{L',L''}(q,\omega)}{d\Omega} = \frac{F_{n0}(q,\omega)}{4\pi}\left\{1 + \frac{6}{W_0}\tilde{G}_{11}\left[\tilde{G}_{00}\cos(\bar{\delta}_0 - \bar{\delta}_1) + 2\tilde{G}_{22}\cos(\bar{\delta}_1 - \bar{\delta}_2)\right]\right.$$
$$P_1(\cos\theta) + \frac{2}{7W_0}\left[21\tilde{G}_{11}^2 + 5\tilde{G}_{22}(7\tilde{G}_{00}\cos(\bar{\delta}_0 - \bar{\delta}_2) + 5\tilde{G}_{22})\right]P_2(\cos\theta) +$$
$$\left.\frac{18}{W_0}\tilde{G}_{11}\tilde{G}_{22}\cos(\bar{\delta}_1 - \bar{\delta}_2)P_3(\cos\theta) + \frac{90}{7W_0}\tilde{G}_{22}^2 P_4(\cos\theta)\right\} \quad (8)$$
$$\equiv \frac{F_{ns}(q,\omega)}{4\pi}\left\{1 - \frac{1}{2}\beta_{ns}^{(in)}(q,\omega)P_2(\cos\theta) + q\left[\gamma_{ns}^{(in)}(q,\omega)P_1(\cos\theta) + \eta_{ns}^{(in)}(q,\omega)P_3(\cos\theta)\right]\right\}$$

where

$$F_{ns} = \frac{4\pi^2 \omega}{q^2} W_0; \quad W_0 = \tilde{G}_{00}^2 + 3\tilde{G}_{11}^2 + 5\tilde{G}_{22}^2. \tag{9}$$

We will compare the result obtained in the small $q$ limit with the known formula for photoionization of an atom by non-polarized light. To do this, we have to use the lowest order terms of the first three spherical Bessel functions:

$$j_0(qr) \cong 1 - \frac{(qr)^2}{6}; \quad j_1(qr) \cong \frac{qr}{3}\left(1 - \frac{(qr)^2}{10}\right); \quad j_2(qr) \cong \frac{(qr)^2}{15}\left(1 - \frac{(qr)^2}{14}\right); \quad j_3(qr) \cong \frac{(qr)^3}{105}. \tag{10}$$

The lowest in powers of $q$ term is $\tilde{G}_{11} \sim q \ll 1$[2]. Correction to $\tilde{G}_{11}$ is proportional to $q^3$. As to $\tilde{G}_{00}$ and $\tilde{G}_{22}$, they are proportional to $q^2$ with corrections of the order of $q^4$. By retaining in (8) terms of the order of $q^2$ and bigger, one has the following expression:

$$\frac{dF_{n0}(q,\omega)}{d\Omega} = \frac{F_{ns}(q,\omega)}{4\pi}\left\{1 + 2P_2(\cos\theta) + \frac{2}{\tilde{G}_{11}}\left[\tilde{G}_{00}\cos(\bar{\delta}_0 - \bar{\delta}_1) + 2\tilde{G}_{22}\cos(\bar{\delta}_1 - \bar{\delta}_2)\right]P_1(\cos\theta) + \right.$$
$$\left. \frac{6\tilde{G}_{22}}{\tilde{G}_{11}}\cos(\bar{\delta}_1 - \bar{\delta}_2)P_3(\cos\theta)\right\} \equiv$$
$$\equiv \frac{F_{ns}(q,\omega)}{4\pi}\left\{1 + 2P_2(\cos\theta) + q\left[\gamma_{ns}^{(in)}(q,\omega)P_1(\cos\theta) + \eta_{ns}^{(in)}(q,\omega)P_3(\cos\theta)\right]\right\} \tag{11}$$

One should compare this relation with the similar one for photoionization of $n0$ subshell [16]:

$$\frac{d\sigma_{ns}(\omega)}{d\Omega} = \frac{\sigma_{ns}(\omega)}{4\pi}\left\{1 - P_2(\cos\theta) + \kappa\frac{6\tilde{Q}_2}{5\tilde{D}_1}\cos(\bar{\delta}_1 - \bar{\delta}_2)\left[P_1(\cos\theta) - P_3(\cos\theta)\right]\right\} \equiv$$
$$\equiv \frac{\sigma_{ns}(\omega)}{4\pi}\left\{1 - P_2(\cos\theta) + \kappa\left[\gamma_{ns}(\omega)P_1(\cos\theta) + \eta_{ns}(\omega)P_3(\cos\theta)\right]\right\}. \tag{12}$$

where $\gamma_{n0}(\omega) = -\eta_{n0}(\omega) = \frac{6\tilde{Q}_2}{5\tilde{D}_1}\cos(\bar{\delta}_1 - \bar{\delta}_2)$.

The difference between (11) and (12) is seen in the sign and magnitude of the dipole parameter that is in electron scattering two times bigger than in photoionization and positive and in different expressions for the non-dipole terms. This difference exists and is essential even in the so-called optical limit $q \to 0$. According to (10), there are simple relations in the $q \to 0$ limit between dipole $\tilde{D}_1$ and quadrupole $\tilde{Q}_2$ matrix elements and functions $\tilde{G}_{11}$, $\tilde{G}_{22}$: $\tilde{G}_{11} = q\tilde{D}_1/3$ and $\tilde{G}_{22} = 2q^2\tilde{Q}_2/15$. With the help of relations $\tilde{G}_{00} = -q^2\tilde{Q}_2/3 = -(5/2)\tilde{G}_{22}$, (11) transforms itself into the following expression:

---

[2] As is seen from (10), we have in mind such values of $q$ that it is $qR_{nl} < 1$, where $R_{nl}$ is the radius of the ionized subshell.

$$\frac{dF_{ns}(q,\omega)}{d\Omega} =$$
$$\frac{F_{ns}(q,\omega)}{4\pi}\left\{1+2P_2(\cos\theta)+q\left[\frac{2\tilde{Q}_2}{\tilde{D}_1}\left(\frac{4}{5}\cos(\bar{\delta}_1-\bar{\delta}_2)-\cos(\bar{\delta}_0-\bar{\delta}_1)\right)P_1(\cos\theta)+2\gamma_{ns}(\omega)P_3(\cos\theta)\right]\right\}$$
(13)

The deviation from (12) is evident, since one cannot express the angular distribution via a single non-dipole parameter $\gamma_{n0}(\omega)$ - a new phase difference $\bar{\delta}_0-\bar{\delta}_1$ appears. As a result, the following relations have to be valid at very small $q$:

$$\gamma_{ns}^{(in)}(\omega)=\frac{2\tilde{Q}_2}{\tilde{D}_1}\left[\frac{4}{5}\cos(\bar{\delta}_1-\bar{\delta}_2)-\cos(\bar{\delta}_0-\bar{\delta}_1)\right],$$
$$\eta_{ns}^{(in)}(\omega)=2\gamma_{n0}(\omega)=\frac{12}{5}\frac{\tilde{Q}_2}{\tilde{D}_1}\cos(\bar{\delta}_1-\bar{\delta}_2).$$
(14)

We see that the investigation of inelastic scattering even at $q\to 0$ permits to obtain an additional characteristic of the ionization process, namely, its s-wave phase.

For $l>0$, even at very small $q$, the relations between non-dipole parameters in photoionization and inelastic fast electron scattering are rather complex.

The similarity of general structure and considerable difference between (11) and (12) is evident. Indeed, the contribution of the non-dipole parameters can be enhanced, since the condition $\omega/c \ll q \ll R^{-1}$ is easy to achieve. Let us note that even while neglecting the terms with $q$, (12) and (13) remain different: in photoionization, the angular distribution is proportional to $\sin^2\theta$ (see (12)), whereas in inelastic scattering it is proportional to $\cos^2\theta$ (see (13)). The reason for this difference is clear. In photoabsorption, the atomic electron is "pushed" off the atom by the electric field of the photon, which is perpendicular to the direction of the light beam. In inelastic scattering, the push acts along momentum $\vec{q}$, so the preferential emission of the electrons takes place along the $\vec{q}$ direction, so the maximum is at $\theta=0$. Similar reason explains the difference in the non-dipole terms. Note that the last term due to monopole transition in (13) is absent in photoabsorption angular distribution (12). It confirms that the angular distribution of the GOS densities is richer than that of photoionization.

The expressions for *p*- and *d*-subshells are much more complex that for s. On the other hand, they are of greater importance and interest since the respective outer *p*- and *d*– subshells are much bigger than that of the s-shell. Multi-electron effects in 4*d*- subshell are particularly important due to presence of the famous dipole Giant resonance. This is why it is of interest to present data on non-s-subshells.

For differential GOSes of *p*-subshells, $l=1$, the following expression is obtained:

$$\frac{dF_{np}(q,\omega)}{d\Omega} = \sum_{L',L''=0}^{2} \frac{dF_{np}^{L',L''}(q,\omega)}{d\Omega} = \frac{F_{np}}{4\pi}\{1+$$

$$\frac{1}{5W_1}\Big[10\tilde{G}_{01}(2\tilde{G}_{12}-\tilde{G}_{10})\cos(\bar{\delta}_0-\bar{\delta}_1) + 4\tilde{G}_{21}\big((5\tilde{G}_{10}-\tilde{G}_{12})\cos(\bar{\delta}_1-\bar{\delta}_2) + 9\bar{G}_{32}\cos(\bar{\delta}_2-\bar{\delta}_3)\big)\Big]P_1(\cos\theta)+$$

$$\frac{2}{7W_1}\Big[7\tilde{G}_{21}(\tilde{G}_{21}-2\tilde{G}_{01}\cos(\bar{\delta}_0-\bar{\delta}_2)) + 7\tilde{G}_{12}(\tilde{G}_{12}-2\tilde{G}_{10})+$$

$$3\tilde{G}_{32}((7\tilde{G}_{10}-2\tilde{G}_{12})\cos(\bar{\delta}_1-\bar{\delta}_3)+4\tilde{G}_{32})\Big]P_2(\cos\theta)-$$

$$\frac{6}{5W_1}\Big[6\tilde{G}_{21}\tilde{G}_{12}\cos(\bar{\delta}_1-\bar{\delta}_2) + \tilde{G}_{32}(5\tilde{G}_{01}\cos(\bar{\delta}_0-\bar{\delta}_3) - 4\tilde{G}_{21}\cos(\bar{\delta}_2-\bar{\delta}_3))\Big]P_3(\cos\theta)+$$

$$\frac{18}{7W_1}\tilde{G}_{32}\Big[\tilde{G}_{32}-4\tilde{G}_{12}\cos(\bar{\delta}_1-\bar{\delta}_3)\Big]P_4(\cos\theta)\} \equiv \frac{F_{np}(q,\omega)}{4\pi}\times$$

$$\left\{1-\frac{\beta_{np}^{(in)}(q,\omega)}{2}P_2(\cos\theta) + q\Big[\gamma_{np}^{(in)}(q,\omega)P_1(\cos\theta) + \eta_{np}^{(in)}(q,\omega)P_3(\cos\theta) + \zeta_{np}^{(in)}(q,\omega)P_4(\cos\theta)\Big]\right\}$$

(15)

where

$$F_{np} = \frac{4\pi^2\omega}{q^2}W_1;\quad W_1 = \tilde{G}_{10}^2 + \tilde{G}_{01}^2 + 2\Big[\tilde{G}_{21}^2 + \tilde{G}_{12}^2\Big] + 3\tilde{G}_{32}^2. \tag{16}$$

For differential GOSes of $d$-subshells, $l=2$, the following expression is obtained:

$$\frac{dF_{nd}(q,\omega)}{d\Omega} = \frac{F_{nd}}{4\pi}\{1 + \frac{6}{W_2}\Big[14\tilde{G}_{11}(\tilde{G}_{22}-\tilde{G}_{20})\cos(\bar{\delta}_1-\bar{\delta}_2) - 14\tilde{G}_{11}\tilde{G}_{02}\cos(\bar{\delta}_0-\bar{\delta}_1)+$$

$$3\tilde{G}_{31}\big((7\tilde{G}_{20}-2\tilde{G}_{22})\cos(\bar{\delta}_2-\bar{\delta}_3)+12\tilde{G}_{42}\cos(\bar{\delta}_3-\bar{\delta}_4)\big)\Big]P_1(\cos\theta)+$$

$$\frac{2}{245W_2}\Big[1029(\tilde{G}_{11}^2+6\tilde{G}_{31}^2) - 18522\tilde{G}_{11}\tilde{G}_{31}\cos(\bar{\delta}_1-\bar{\delta}_3) + 1225\tilde{G}_{02}(7\tilde{G}_{20}-10\tilde{G}_{22})\cos(\bar{\delta}_0-\bar{\delta}_2)-$$

$$125\tilde{G}_{22}(98\tilde{G}_{20}+15\tilde{G}_{22}) + 450\tilde{G}_{42}((49\tilde{G}_{20}-20\tilde{G}_{22})\cos(\bar{\delta}_2-\bar{\delta}_4)+25\tilde{G}_{42})\Big]P_2(\cos\theta)+$$

$$\frac{18}{W_2}\Big[2\tilde{G}_{11}(\tilde{G}_{22}\cos(\bar{\delta}_1-\bar{\delta}_2)-6\tilde{G}_{42}\cos(\bar{\delta}_1-\bar{\delta}_4)) + \tilde{G}_{31}(7\tilde{G}_{02}\cos(\bar{\delta}_0-\bar{\delta}_3)-$$

$$8\tilde{G}_{22}\cos(\bar{\delta}_2-\bar{\delta}_3)+6\tilde{G}_{42}\cos(\bar{\delta}_3-\bar{\delta}_4))\Big]P_3(\cos\theta)+$$

$$\frac{90}{49W_2}\Big[20\tilde{G}_{22}^2 + \tilde{G}_{42}(98\tilde{G}_{02}\cos(\bar{\delta}_0-\bar{\delta}_4)-100\tilde{G}_{22}\cos(\bar{\delta}_2-\bar{\delta}_4)+27\tilde{G}_{42})\Big]P_4(\cos\theta)\} \equiv$$

$$\frac{F_{nd}(q,\omega)}{4\pi}\times$$

$$\left\{1-\frac{\beta_{nd}^{(in)}(q,\omega)}{2}P_2(\cos\theta) + q\Big[\gamma_{nd}^{(in)}(q,\omega)P_1(\cos\theta) + \eta_{nd}^{(in)}(q,\omega)P_3(\cos\theta) + \zeta_{nd}^{(in)}(q,\omega)P_4(\cos\theta)\Big]\right\}, (17)$$

where

$$F_{nd} = \frac{4\pi^2\omega}{35q^2}W_2;\quad W_2 = 35\tilde{G}_{20}^2 + 42\tilde{G}_{11}^2 + 63\tilde{G}_{31}^2 + 35\tilde{G}_{02}^2 + 50\tilde{G}_{22}^2 + 90\tilde{G}_{42}^2. \tag{18}$$

Of interest is to compare, just as was done with $l=0$, the expressions (15) and (17) with angular distribution of photoelectrons. It is essential to clarify whether the difference exists even in the $q \to 0$ limit, as it takes place for the $s$- subshells. In this limit the following expressions follow from (15) and (17):

For $l=1$ one has from (15) at $q=0$

$$\beta_{np}^{(in)}(q=0,\omega) = -\frac{4}{\tilde{D}_0^2 + 2\tilde{D}_2^2}[\tilde{D}_2^2 - 2\tilde{D}_0\tilde{D}_2 \cos(\bar{\delta}_0 - \bar{\delta}_2)], \qquad (19)$$

$$\gamma_{np}^{(in)}(q=0,\omega) = \frac{18}{25[\tilde{D}_0^2 + 2\tilde{D}_2^2]}\{5\tilde{D}_0\tilde{Q}_1 \cos(\bar{\delta}_1 - \bar{\delta}_0) + 2\tilde{D}_2[2\tilde{Q}_3 \cos(\bar{\delta}_3 - \bar{\delta}_2) - 3\tilde{Q}_1 \cos(\bar{\delta}_1 - \bar{\delta}_2)\}, \qquad (20)$$

$$\eta_{np}^{(in)}(q=0,\omega) = \frac{12}{25[\tilde{D}_0^2 + 2\tilde{D}_2^2]}\{5\tilde{D}_0\tilde{Q}_3 \cos(\bar{\delta}_3 - \bar{\delta}_0) + 2\tilde{D}_2\left[3\tilde{Q}_1 \cos(\bar{\delta}_1 - \bar{\delta}_2) - 2\tilde{Q}_3 \cos(\bar{\delta}_3 - \bar{\delta}_2)\right]\} \qquad (21)$$

For $l=2$ one has from (17) at $q=0$

$$\beta_{nd}^{(in)}(q=0,\omega) = -\frac{4}{5[2\tilde{D}_1^2 + 3\tilde{D}_3^2]}[\tilde{D}_1^2 + 6\tilde{D}_3^2 - 18\tilde{D}_1\tilde{D}_3 \cos(\bar{\delta}_1 - \bar{\delta}_3)], \qquad (22)$$

$$\gamma_{nd}^{(in)}(q=0,\omega) = \frac{2}{35[2\tilde{D}_1^2 + 3\tilde{D}_3^2]} \times$$
$$\{14\tilde{D}_1\left[7\tilde{Q}_2 \cos(\bar{\delta}_1 - \bar{\delta}_2) - 2\tilde{Q}_0 \cos(\bar{\delta}_0 - \bar{\delta}_1)\right] + 9\tilde{D}_3\left[8\tilde{Q}_4 \cos(\bar{\delta}_4 - \bar{\delta}_3) - 13\tilde{Q}_2 \cos(\bar{\delta}_2 - \bar{\delta}_3)\right]\}, \qquad (23)$$

$$\eta_{nd}^{(in)}(q=0,\omega) = \frac{12}{35[2\tilde{D}_1^2 + 3\tilde{D}_3^2]}\{2\tilde{D}_1\left[\tilde{Q}_2 \cos(\bar{\delta}_2 - \bar{\delta}_1) - 6\tilde{Q}_4 \cos(\bar{\delta}_4 - \bar{\delta}_1)\right] +$$
$$\tilde{D}_3\left[7\tilde{Q}_0 \cos(\bar{\delta}_0 - \bar{\delta}_3) - 8\tilde{Q}_2 \cos(\bar{\delta}_2 - \bar{\delta}_3) - 6\tilde{Q}_4 \cos(\bar{\delta}_4 - \bar{\delta}_3)\right]\} \qquad (24)$$

In deriving, (19-24) the following relations were used

$$\tilde{G}_{l'1} \equiv \frac{q}{3}\tilde{D}_{l'}(l'=l\pm 1); \tilde{G}_{l'0} \equiv -\frac{q^2}{3}\tilde{Q}_{l'}(l'=l); \tilde{G}_{l'2} \equiv \frac{2q^2}{15}\tilde{Q}_{l'}(l'=l,l\pm 2); \qquad (25)$$

To clarify comparison between angular anisotropy parameters in photoionization and fast electron scattering, note that the following relations are used in the HF approximation

$$\tilde{D}_{l'} \Rightarrow \tilde{d}_{l'} = \int_0^\infty P_{nl}(r) r P_{kl'}(r) dr; \; \tilde{Q}_{l'} \Rightarrow \tilde{q}_{l'} = \frac{1}{2}\int_0^\infty P_{nl}(r) r^2 P_{kl'}(r) dr \;, \qquad (26)$$

where $P_{nl(kl')}(r) = rR_{nl(kl')}(r)$ and $R_{nl(kl')}(r)$ are the radial parts of the HF electron wave functions in the initial (final) states.

The angular distribution of photoelectrons with inclusion of the lowest order in photon momentum $\kappa = \omega/c$ non-dipole terms is given by the following expression for any $l$:

$$\frac{d\sigma_{nl}(\omega)}{d\Omega} = \frac{\sigma_{nl}(\omega)}{4\pi}\left\{1 - \frac{\beta_{nl}(\omega)}{2}P_2(\cos\theta) + \kappa\gamma_{nl}(\omega)P_1(\cos\theta) + \kappa\eta_{nl}(\omega)P_3(\cos\theta)\right\}. \quad (27)$$

For $l=1$ one has the following expression for the dipole angular anisotropy parameters [10, 1]

$$\beta_{np}(\omega) = \frac{2}{\tilde{D}_0^2 + 2\tilde{D}_2^2}[\tilde{D}_2^2 - 2\tilde{D}_0\tilde{D}_2\cos(\bar{\delta}_0 - \bar{\delta}_2)]. \quad (28)$$

As is seen from (19), for $l=1$ the relation $\beta_{np}^{(in)}(q=0,\omega) = -2\beta_{np}(\omega)$ is the same as for the $s$-subshells.

The following expressions determine the non-dipole angular anisotropy parameters [1] for $l=1$:

$$\gamma_{np}(\omega) = \frac{6}{25[\tilde{D}_0^2 + 2\tilde{D}_2^2]}\left\{5\tilde{D}_0\tilde{Q}_1\cos(\bar{\delta}_1 - \bar{\delta}_0) + \tilde{D}_2\left[9\tilde{Q}_3\cos(\bar{\delta}_3 - \bar{\delta}_2) - \tilde{Q}_1\cos(\bar{\delta}_1 - \bar{\delta}_2)\right]\right\}, \quad (29)$$

$$\eta_{np}(\omega) = \frac{6}{25[\tilde{D}_0^2 + 2\tilde{D}_2^2]}\left\{5\tilde{D}_0\tilde{Q}_3\cos(\bar{\delta}_3 - \bar{\delta}_0) + 2\tilde{D}_2\left[3\tilde{Q}_1\cos(\bar{\delta}_1 - \bar{\delta}_2) - 2\tilde{Q}_3\cos(\bar{\delta}_3 - \bar{\delta}_2)\right]\right\}. \quad (30)$$

One has for the dipole angular anisotropy parameter the following expression [10] for $l=2$:

$$\beta_{nd}(\omega) = \frac{2}{5[2\tilde{D}_1^2 + 3\tilde{D}_3^2]}[\tilde{D}_1^2 + 6\tilde{D}_3^2 - 18\tilde{D}_1\tilde{D}_3\cos(\bar{\delta}_1 - \bar{\delta}_3)], \quad (31)$$

Note that as is seen from (22), for $l=2$ the relation similar to $l=0;1$ is valid: $\beta_{nd}^{(in)}(q=0,\omega) = -2\beta_{nd}(\omega)$. Quite possible that such a relation is valid for any $l$.

The following expressions determine the non-dipole angular anisotropy parameters [1] for $l=2$

$$\gamma_{nd}(\omega) = \frac{6}{35[2\tilde{D}_1^2 + 3\tilde{D}_3^2]}\left\{7\tilde{D}_1[\tilde{Q}_2\cos(\bar{\delta}_2 - \bar{\delta}_1) - \tilde{Q}_0\cos(\bar{\delta}_0 - \bar{\delta}_1)] + \right.$$
$$\left. 3\tilde{D}_3[6\tilde{Q}_4\cos(\bar{\delta}_4 - \bar{\delta}_3) - \tilde{Q}_2\cos(\bar{\delta}_2 - \bar{\delta}_3)]\right\} \quad (32)$$

$$\eta_{nd}(\omega) = \frac{6}{35[2\tilde{D}_1^2 + 3\tilde{D}_3^2]}\left\{2\tilde{D}_1\left[6\tilde{Q}_4\cos(\bar{\delta}_4 - \bar{\delta}_1) - \tilde{Q}_2\cos(\bar{\delta}_2 - \bar{\delta}_1)\right] - \right.$$
$$\left. \tilde{D}_3\left[8\tilde{Q}_2\cos(\bar{\delta}_2 - \bar{\delta}_3) - 6\tilde{Q}_4\cos(\bar{\delta}_4 - \bar{\delta}_3) - 7\tilde{Q}_0\cos(\bar{\delta}_0 - \bar{\delta}_3)\right]\right\} \quad (33)$$

Prominent analytic deviation is seen from respective non-dipole parameters for inelastic scattering, given by (20, 21, 23, 24). Contrary to the dipole parameters, simple frequency independent relations that connect respective non-dipole parameters for photoionization and fast electron inelastic scattering do not exist.

Note that the limit $q=0$ at $\omega \neq 0$ cannot be achieved since no energy can be transferred from the incoming electron to the projectile without momentum transfer. However, with growth of the projectile's speed, smaller and smaller $q$ is sufficient to transfer given energy $\omega$.

In spite of visibly deep similarity between photoionization and fast electron scattering, as is seen, a big difference exists. Indeed, the angular distributions in photoionization and fast electron scattering are different even in the limit $q \to 0$. It can be explained by the difference between a transverse (in photoionization) and longitudinal (in fast electron scattering) photons that ionize the target atom. Analytic, it is reflected in the difference between operators causing ionization by photons and fast electrons that include already only the lowest non-dipole corrections. For photoionization it is $(\vec{e}\vec{r}) + i(\vec{\kappa}\vec{r})(\vec{e}\vec{r})$, where $\vec{e}$ is the photon polarization operator that is orthogonal to the direction of light propagation. As to fast electron scattering, it is $(\vec{q}\vec{r}) + i(\vec{q}\vec{r})(\vec{q}\vec{r})$, thus including only one angle between $\vec{q}$ and $\vec{r}$ contrary to the case of photoionization with its two angles – between $\vec{r}$, $\vec{e}$ and $\vec{\kappa}$.

Because of this difference, in photoionization the force that acts upon the outgoing electron is orthogonal to the direction of photon momentum $\vec{\kappa}/\kappa$ and thus of the photon beam. Therefore the photoelectron emission is minimal along $\vec{\kappa}/\kappa$, while in inelastic electron scattering the force and maximal knocked-out electron yield is directed along $\vec{q}$.

## 3. Calculation details

In order to obtain $dF_{nl}(q,\omega)/d\Omega$ from experiment, one has to measure the yield of electrons emitted at a given angle $\theta$ with energy $\varepsilon = k^2/2 = \omega - I_{nl}$ in coincidence with the fast outdoing particle that looses energy $\omega$ and transfers to the target atom momentum $\vec{q}$. Note that $\beta_{n0}^{(in)}$ is (-4) that differs by sign and value from photoionization value $\beta_{ns} = 2$. As we will show, essential are the differences for other considered subshells.

To calculate $dF_{nl}(q,\omega)/d\Omega$ we used the numeric procedures described at length in [14]. Calculations are performed in the frame of Hartree-Fock and RPAE approximations. As concrete objects, we choose outer and subvalent subshells $np^2$ and $ns^2$ subshells of Ar, Kr and Xe. Non-dipole parameter $\zeta_{nd}^{(in)}$ was calculated for 3d Kr and 4d Xe.

We perform calculations using equations (5-9, 11, 13, 15-18) in HF and RPAE, for $q = 0.0$, $0.1$, $1.1$ and $2.1$ at. un. The energies of outgoing electrons is considered up to 20-25 Ry. Note, however that the point $q = 2.1$ is given for some orientation since for not small enough q-values the formula presented and discussed in this paper are incorrect: with growth of q values $L', L'' > 2$ become increasingly important.

Most prominent are the non-dipole corrections at so-called magic angle $\theta_m$, at which the following relation holds: $P_2(\cos\theta_m) = 0$. This is why differential in emission angle GOSes $dF_{nl}(q,\omega)/d\Omega$ are presented at the magic angle $\theta_m$ and at $q = 0.0; 0.1; 1.1; 2.1$. Results are obtained also for dipole and non-dipole angular anisotropy parameters. All data are presented in Fig.1-11.

The lowest value of $q$ corresponds to the photoionization limit, since $qR \ll 1$ and in the considered frequency range $\omega/c < 0.05 < q_{min} = 0.1$. The last inequality shows that we consider non-dipole corrections to the GOSes that are much bigger than the non-dipole corrections to photoionization.

## 4. Calculation results

The results demonstrate that the GOSes and angular anisotropy parameters are complex and informative functions with a number of prominent variations. All calculated characteristics demonstrate strong influence of the electron correlations for *p*-, *s*-, and *d*- electrons. They depend strongly upon the outgoing electron energy and the linear momentum *q* transferred to the atom in fast electron inelastic scattering, being strongly affected by electron correlations.

In Fig. 1 and 2 differential generalized oscillator strengths (GOSes) given by (8) and (15) at the magic angle $P_2(\cos\theta_m) = 0$, $\theta_m \cong 54.7^0$ for outer *np*- and subvalent *ns*-subshells of Ar, Kr and Xe at q=0.1, 1.1 and 2.1 are presented in HF and RPAE. At small q the GOSes are similar to the photoionization cross-section. For p-subshells with growth of q the maximum decreases in magnitude and shifts to higher $\omega$. For q=2.1 there is no traces of any similarity with photoionization. The situation for s-subshell is different, since there the differential GOSes with increase of q at first grow and then start rapidly to decrease.

The insertion in Fig.1 for 5p Xe shows the prominent effect played by the action of 4d Giant resonance upon 5p GOS. Note that for big q, q=2.1 the maximum exists at the same energy already in HF, and the action of 4d adds only a small shoulder.

Fig. 3-5 collects the non-dipole angular anisotropy of knocked-out electrons $\gamma_{ns}^{(in)}(\omega)$ and $\eta_{ns}^{(in)}(\omega)$ given by (7) and (14) at q=0.01, 0.1, and 1.1, compared to similar parameters in photoionization $\gamma_{ns}(\omega)$ and $\eta_{ns}(\omega)$ (11) for subvalent ns subshell of Ar, Kr and Xe in RPAE. For q=0 the relation $\eta_{ns}^{(in)}(\omega) = 2\gamma_{n0}(\omega)$ is valid. As to $\gamma_{ns}^{(in)}(\omega)$, it is of different sign and three-to-four times bigger than $\eta_{ns}(\omega)$. It means that even in the limit q=0 the non-dipole parameters for photoionization and for fast electron inelastic scattering are essentially different. Qualitatively, parameters at q=0.1 look similar to that at q=0, but smaller. With increase of q the variation become broader and shifted to the higher $\omega$ side. Note that an approximate relation proved to be valid between $\gamma_{ns}^{(in)}(\omega)$ and $\eta_{ns}^{(in)}(\omega)$.

Fig. 6 presents the dipole angular anisotropy parameter of knocked-out electrons $\beta_{np}^{(in)}(q,\omega)$ given by (15) at q=0.1 and 1.1, compared to similar parameter in photoionization $\beta_{np}(\omega)$, given by (28) for outer subshells of Ar, Kr and Xe in RPAE. It is seen that for q=0.1 the relation, that is precisely correct at q=0, $\beta_{np}^{(in)}(q=0,\omega) = -2\beta_{np}(\omega)$, is approximately valid, while it is violated for bigger q. It looks like the following relation is valid $-4 \leq \beta_{nl}^{(in)}(q,\omega) \leq 2$. Maximum for $\beta_{np}(\omega)$ and minima for $\beta_{np}^{(in)}(q,\omega)$ in Xe in the $\omega$ region around 8-10 Ry are consequences of the effect of the 4d Giant resonance.

Fig. 7 depicts the angular anisotropy non-dipole parameters of knocked-out electrons $\gamma_{np}^{(in)}(q,\omega)$ and $\eta_{np}^{(in)}(q,\omega)$ given by (15) at q=0.1 and 1.1, compared to similar parameters in photoionization $\gamma_{np}(\omega)$ and $\eta_{np}(\omega)$ given by (29, 30) for 3p Ar, 4p Kr and 5p Xe subshells in RPAE. As is already seen from the analytic expressions, the difference between photoionization values and that for fast electron scattering is essential even in the limit q=0. The non-dipole parameters are complex and thus rather informative functions of $\omega$ at both q-values.

Fig.8 represents the angular anisotropy dipole parameter of knocked-out electrons $\beta_{nd}^{(in)}(q,\omega)$ given by (17) and (22) at q=0.1, 1.1 and 2.1, compared to similar parameters in photoionization $\beta_{nd}(\omega)$, given by (31) for 3d Kr and 4d Xe subshells in RPAE. Note that the relation $\beta_{nd}^{(in)}(q=0,\omega) = -2\beta_{nd}(\omega)$ is fulfilled. Prominent changes of $\beta_{nd}^{(in)}(q,\omega)$ take place with weakening of variations with increase of q.

Fig.9 shows the angular anisotropy non-dipole parameters of knocked-out electrons $\gamma_{nd}^{(in)}(q,\omega)$ and $\eta_{nd}^{(in)}(q,\omega)$ given by (17) at q=0.1 and 1.1, compared to similar parameters in photoionization $\gamma_{nd}(\omega)$ and $\eta_{nd}(\omega)$, given by (32, 33) for 3d Kr and 4d Xe subshells in RPAE. The difference between parameters for photoionization and fast electron scattering is quite big. Note that the parameters, as it should be, are smaller than data for respective p-subshells since the radiuses 3d Kr and 4d Xe are smaller than that of 4p Kr and 5p Xe, respectively.

Fig.10 demonstrates the angular anisotropy non-dipole parameter – coefficient of the fourth Legendre polynomial in the angular distribution of the knocked-out electrons $\zeta_{np}^{(in)}(q,\omega)$, calculated using (15). The results are presented for 3p Ar, 4p Kr, 5p Xe subshells at q=0.1, 1.1 and 2.1. This parameter does not have a calculated photoionization analog. The absolute value is much smaller than other non-dipole parameters for the dame subshells.

Fig. 11 gives the data on the angular anisotropy non-dipole parameter $\zeta_{nd}^{(in)}(q,\omega)$ – coefficient of the fourth Legendre polynomial in the angular distribution of the knocked-out electrons, calculated at q=0.1, 1.1 and q=2.1 using (17). The results are presented for 3d Kr and 4d Xe subshells. These results are, as it should be, much smaller than in Fig. 10.

## 5. Concluding remarks

It is not a surprise that GOSes and angular anisotropy parameters depend upon q. What is indeed a surprise s the big difference between the angular anisotropy parameters for fast electron scattering and respective photoionization values. Already from photoionization studies, we know that they are strongly affected by atomic electron correlations. Here we saw that fast electron scattering gives information also on transferred momentum dependences and their interplay with electron correlations.

The biggest unexpected feature of the angular anisotropy for inelastic scattering is that even in the q=0 limit they do not coincide with respective photoionization values, and they are not connected by simple relation similar to that between photoionization cross-section and GOSes. This is a result of different operators for photoionization and fast electron scattering as is discussed at the very end of Section 3.

We expect that this paper will stimulate experimental efforts in not too simple but potentially rather informative studies of the differential cross-section of secondary electrons knocked out off a target atom in fast electron-atom collisions. We understand that such studies require coincidence experiments, in which simultaneously not only the transferred by fast electron energy and momentum is fixed, but momentum, including its direction, of the secondary electron.

Particular and first attention deserves the $q \to 0$ limit. It is seen that different, by sign and value, are already the dipole angular anisotropy parameters. The non-dipole parameters in their turn deviate even qualitatively from their respective photoionization values. It is amazing that in the non-relativistic domain of energies at first glance inessential difference between a virtual and real photon leads to so powerful consequences.

The information that could come from studies of angular distribution of secondary electrons at small $q$ is of great interest and value. Thus, the suggested here experimental studies are desirable.

## Acknowledgements

The authors are grateful for the financial assistance via the Israeli-Russian grant RFBR-MSTI 11-02-92484

## References

1. *Amusia M. Ya., Baltenkov A. S., Chernysheva L.V., Felfli Z., and Msezane A. Z.* Phys. Rev. A. 2001. V.**63**. 052506.
2. *Cooper J. W.* Phys. Rev. A. 1990. V. **42**, P.6942; Phys. Rev. A. 1992. V. **45**, 3362; Phys. Rev. A. 1993. V. **47**, P.1841;
3. *Bechler A. and Pratt R. H.* Phys. Rev. A. 1990. V **42. P.** 6400.

4. *Krassig B., Jung M., Gemmell D.S., Kanter E.P., LeBrun T., Southworth S.H., Young L.* Phys. Rev. Lett. 1995. V. 175. P. 4736-4739.
5. *Kanter E.P., Krassig B., Southworth S.H., Guillemin R., Hemmers O.D. Lindle W., Wehlitz R., Amusia M.Ya, Chernysheva L.V., Martin N.L.S.* Phys. Rev. A. 2003. V. 68. P. 012714-1-10.
6. *Hemmers O., Guillemin R., Kanter E.P., Krassig B.,. Lindle D., Southworth S.H., Wehlitz R., Baker J., Hudson A., Lotrakul M., Rolles D., Stolte W.C., Tran I.C., Wolska A., Yu S.W., Amusia M.Ya., Cheng K.T., Chernysheva L.V., Johnson W.R., Manson S.T.* Phys. Rev. Lett. 2003. V. 91(5). P. 053002/1-4.
7. *Kivimäki A., Hergenhahn U., Kempgens B., Hentges R., Piancastelli M.N., Maier K., Ruedel A., Tulkki J.J., Bradshaw A.M.* Phys. Rev. 2000. V. 63. P. 012716.
8. *El-Sherbini T. M., Van der Wiel M. J.*, Physica, 1972
9. *Amusia M. Ya., Cherepkov N. A*. Case Studies in Atomic Physics. 1975. V. 5. P. 47-179.
10. *Amusia M. Ya.,* Atomic Photoeffect, N. Y.-London, Plenum Press, 1990.
11. *M. Inokuti*, Rev. Mod. Phys, **43**, 297, 1971
12. *L. D. Landau and E. M. Lifshitz*, Quantum Mechanics. Non-Relativistic Theory (Pergamon, Oxford, 1965).
13. *Amusia M. Ya.,* Radiation Physics and Chemistry, **70**, 237-251, 2004.
14. *Amusia M. Ya. and Chernysheva L.V.*, *Computation of Atomic Processes*, Institute of Physics Publishing, Bristol and Philadelphia, 1997
15. http://mathworld.wolfram.com/SphericalHarmonic.html
16. http://mathworld.wolfram.com/SphericalHarmonicAdditionTheorem.html
17. *Amusia M.Ya., Arifov P.U., Baltenkov A.S., Grinberg A.A., Shapiro S.G.* Phys. Lett A. 1974, **47**, 66-69

**Figure captions**

Fig. 1. Differential generalized oscillator strength given by (8) at the magic angle $P_2(\cos\theta_m) = 0$, $\theta_m \cong 54.7^0$ of 3p-, 4p-and 5p subshells for Ar, Kr and Xe at q=0.1, 1.1, 2.1 in HF and RPAE.

Fig. 2. Differential generalized oscillator strength given by (15) at the magic angle $P_2(\cos\theta_m) = 0$, $\theta_m \cong 54.7^0$ of 3s-, 4s-and 5s subshells for Ar, Kr and Xe at q=0.1, 1.1, 2.1 in HF and RPAE.

Fig. 3. Angular anisotropy non-dipole parameters of knocked-out electrons $\gamma_{ns}^{(in)}(\omega)$ and $\eta_{ns}^{(in)}(\omega)$ given by (8) and (13) at q=0.01, 0.1, and 1.1, compared to similar parameters in photoionization $\gamma_{ns}(\omega)$ and $\eta_{ns}(\omega)$ (12) for 3s subshell of Ar in RPAE.

Fig. 4. Angular anisotropy non-dipole parameters of knocked-out electrons $\gamma_{ns}^{(in)}(\omega)$ and $\eta_{ns}^{(in)}(\omega)$ given by (8) and (13) at q=0.01, 0.1 and 1.1, compared to similar parameters in photoionization $\gamma_{ns}(\omega)$ and $\eta_{ns}(\omega)$ (12) for 4s subshell of Kr in RPAE.

Fig. 5. Angular anisotropy non-dipole parameters of knocked-out electrons $\gamma_{ns}^{(in)}(\omega)$ and $\eta_{ns}^{(in)}(\omega)$ given by (8) and (13) at q=0.01, 0.1 and 1.1, compared to similar parameters in photoionization $\gamma_{ns}(\omega)$ and $\eta_{ns}(\omega)$ (12) for 5s subshell of Xe in RPAE.

Fig. 6. Angular anisotropy dipole parameter of knocked-out electrons $\beta_{np}^{(in)}(q,\omega)$ given by (15) at q=0.1 and 1.1, compared to similar parameter in photoionization $\beta_{np}(\omega)$, given by (28) for outer subshells of Ar, Kr and Xe in RPAE.

Fig. 7. Angular anisotropy parameters of knocked-out electrons $\gamma_{np}^{(in)}(q,\omega)$ and $\eta_{np}^{(in)}(q,\omega)$ given by (15) at q=0.1 and 1.1, compared to similar parameters in photoionization $\gamma_{np}(\omega)$ and $\eta_{np}(\omega)$ given by (29, 30) for 3p Ar, 4p Kr and 5p Xe subshells in RPAE.

Fig. 8. Angular anisotropy dipole parameter of knocked-out electrons $\beta_{nd}^{(in)}(q,\omega)$ given by (17) and (22) at q=0.1 and 1.1, compared to similar parameters in photoionization $\beta_{nd}(\omega)$, given by (31) for 3d Kr and 4d Xe subshells in RPAE.

Fig.9. Angular anisotropy non-dipole parameters of knocked-out electrons $\gamma_{nd}^{(in)}(q,\omega)$ and $\eta_{nd}^{(in)}(q,\omega)$ given by (17) at q=0.1 and 1.1, compared to similar parameters in photoionization $\gamma_{np}(\omega)$ and $\eta_{np}(\omega)$, given by (29, 30) for 3d Kr and 4d Xe subshells in RPAE.

Fig.10. Angular anisotropy non-dipole parameter – coefficient of the fourth Legendre polynomial in the angular distribution of the knocked-out electrons $\zeta_{np}^{(in)}(q,\omega)$, calculated using (15). The results are presented for 3p Ar, 4p Kr, 5p Xe subshells at q=0.1, 1.1 and 2.1.

Fig. 11. Anisotropy non-dipole parameter $\zeta_{nd}^{(in)}(q,\omega)$ – coefficient of the fourth Legendre polynomial in the angular distribution of the knocked-out electrons $\zeta_{nl}^{(in)}(q,\omega)$, calculated at q=1.1 and q=2.1 using (17). The results are presented for outer and nd subshells of Ar, Kr, and Xe subshells.

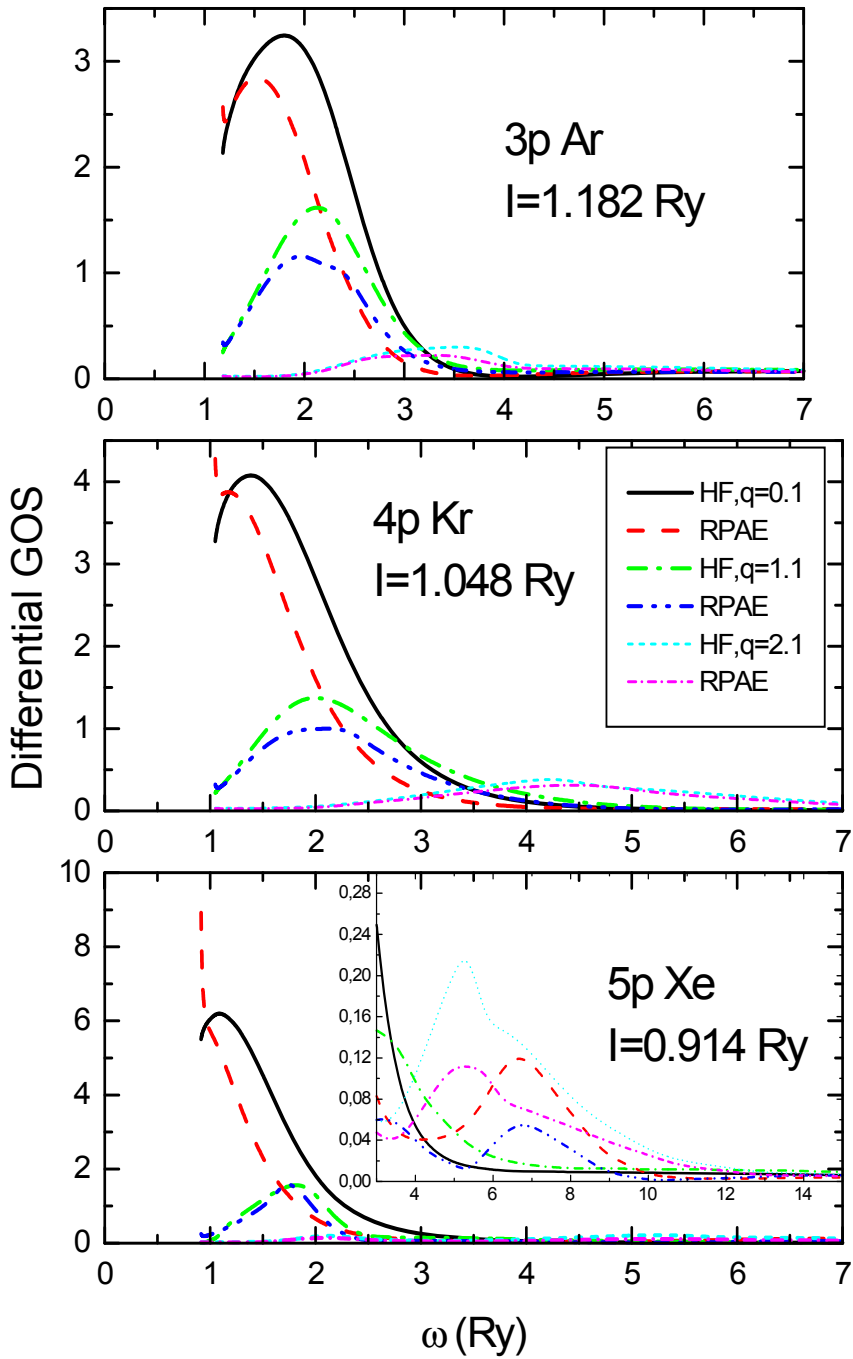

Fig.1 Differential generalized oscillator strength given by (8) at the magic angle $P_2(\cos\theta_m)=0$, $\theta_m \cong 54.7^0$ of 3p-, 4p- and 5p subshells for Ar, Kr and Xe at q=0.1, 1.1, 2.1 in HF and RPAE.

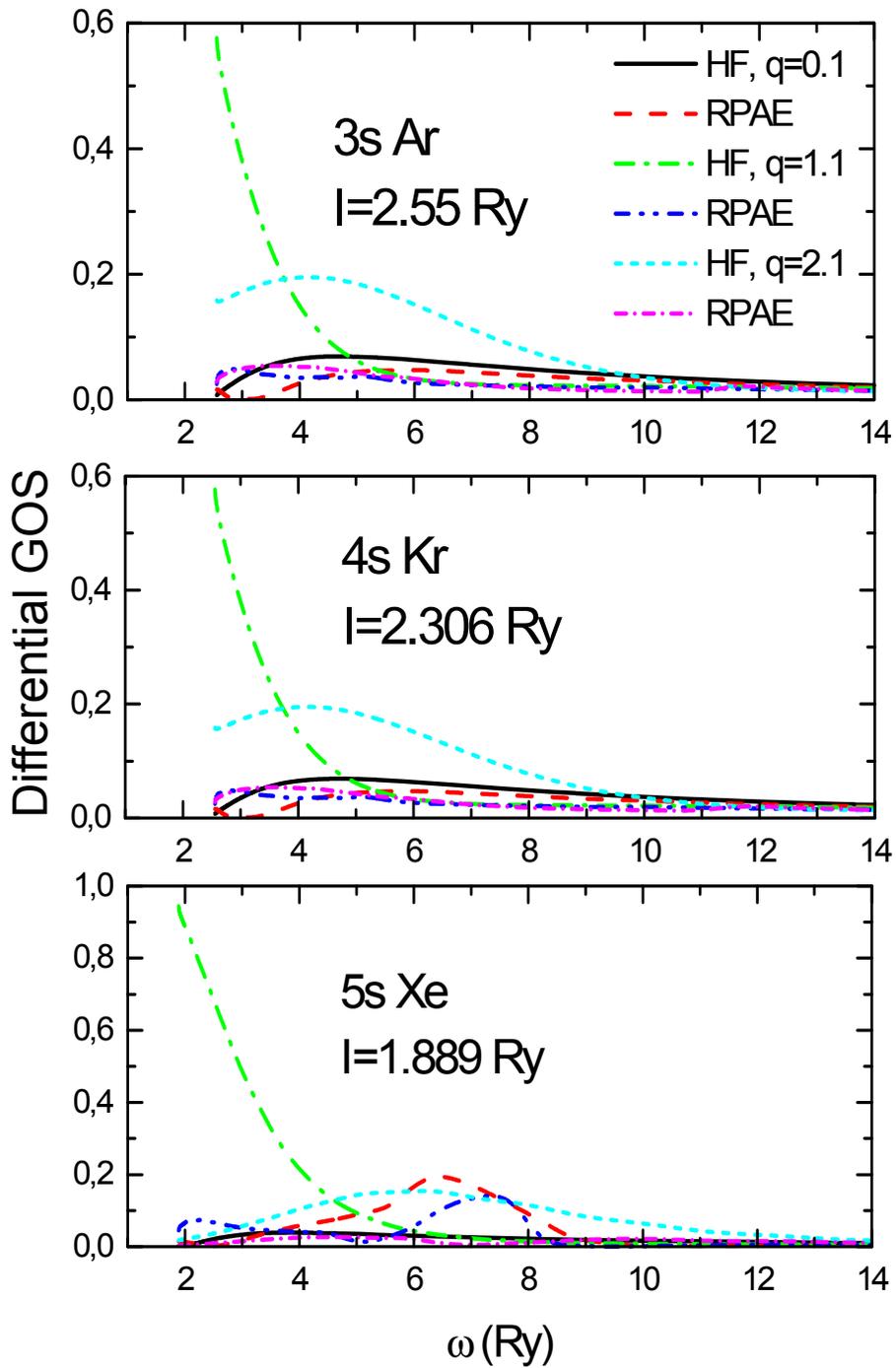

Fig.2. Differential generalized oscillator strength given by (8) at the magic angle $P_2(\cos\theta_m) = 0$, $\theta_m \cong 54.7^0$ of 3s-, 4s- and 5s subshells for Ar, Kr and Xe at q=0.1, 1.1, 2.1 in HF and RPAE.

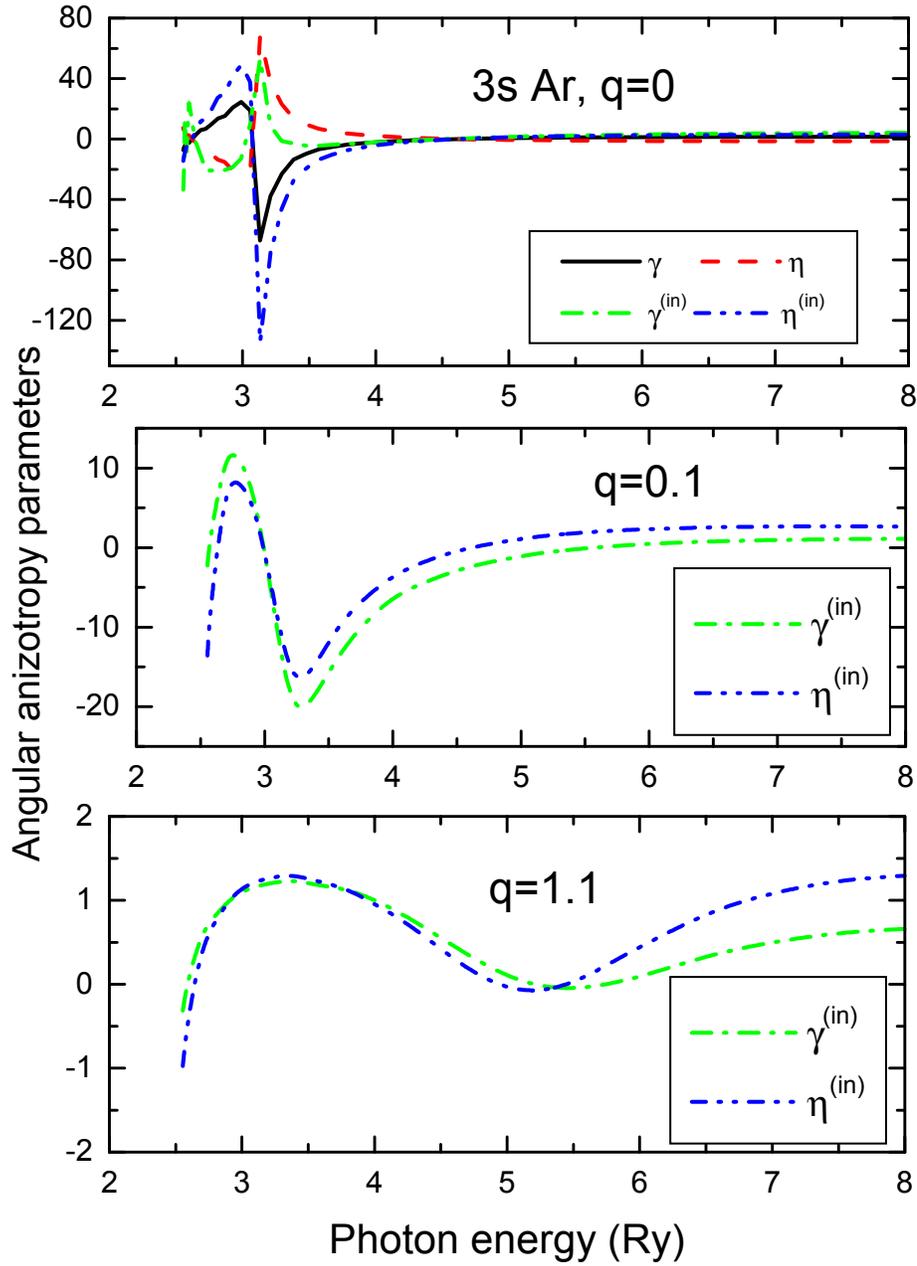

Fig.3. Angular anisotropy non-dipole parameters of knocked-out electrons $\gamma_{ns}^{(in)}(\omega)$ and $\eta_{ns}^{(in)}(\omega)$ given by (7) and (8) at q=0.01, 0.1 and 1.1, compared to similar parameters in photoionization $\gamma_{ns}(\omega)$ and $\eta_{ns}(\omega)$ (12) for 3s subshell of Ar in RPAE

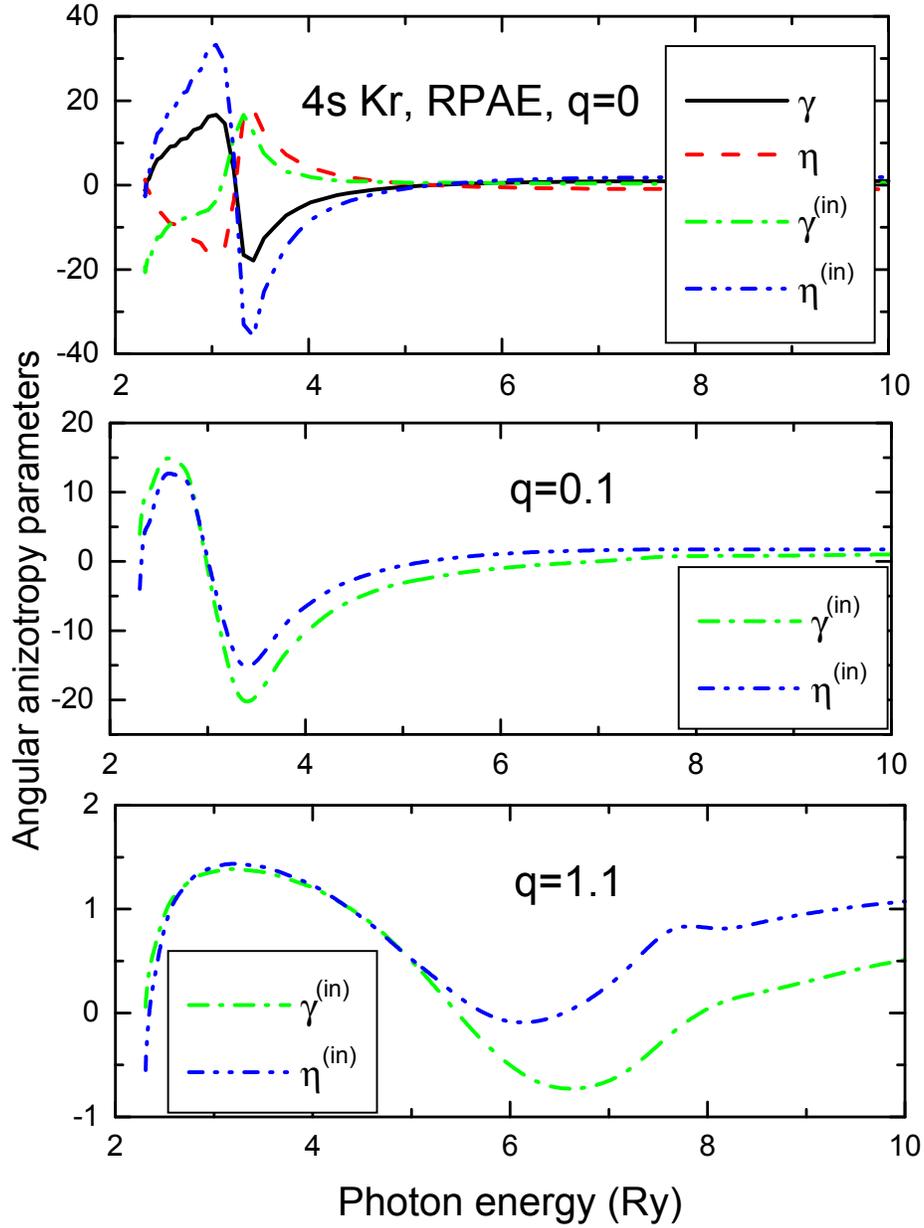

Fig.4. Angular anisotropy non-dipole parameters of knocked-out electrons $\gamma_{ns}^{(in)}(\omega)$ and $\eta_{ns}^{(in)}(\omega)$ given by (7) and (8) at q=0.01, 0.1 and 1.1, compared to similar parameters in photoionization $\gamma_{ns}(\omega)$ and $\eta_{ns}(\omega)$ (12) for 4s subshell of Kr in RPAE.

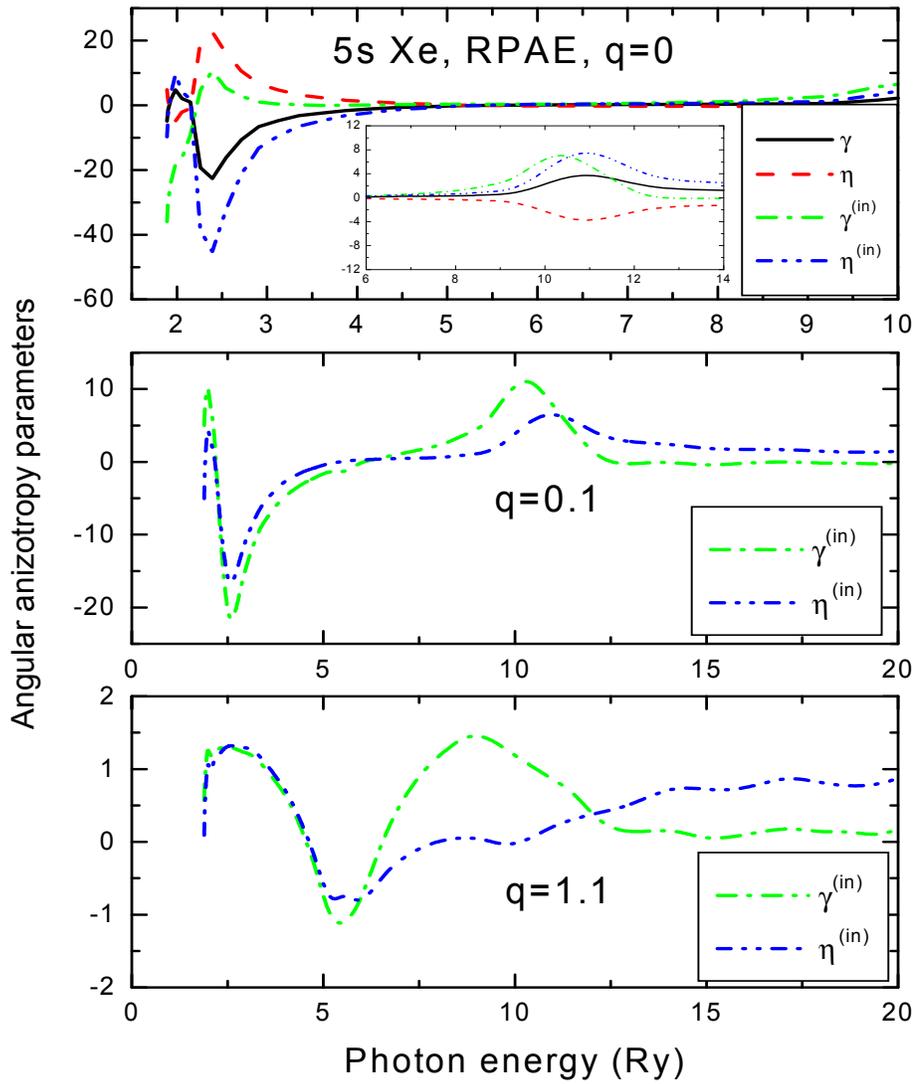

Fig. 5. Angular anisotropy non-dipole parameters of knocked-out electrons $\gamma_{ns}^{(in)}(\omega)$ and $\eta_{ns}^{(in)}(\omega)$ given by (7) and (8) at q=0.01, 0.1 and 1.1, compared to similar parameters in photoionization $\gamma_{ns}(\omega)$ and $\eta_{ns}(\omega)$ (12) for 5s subshell of Xe in RPAE.

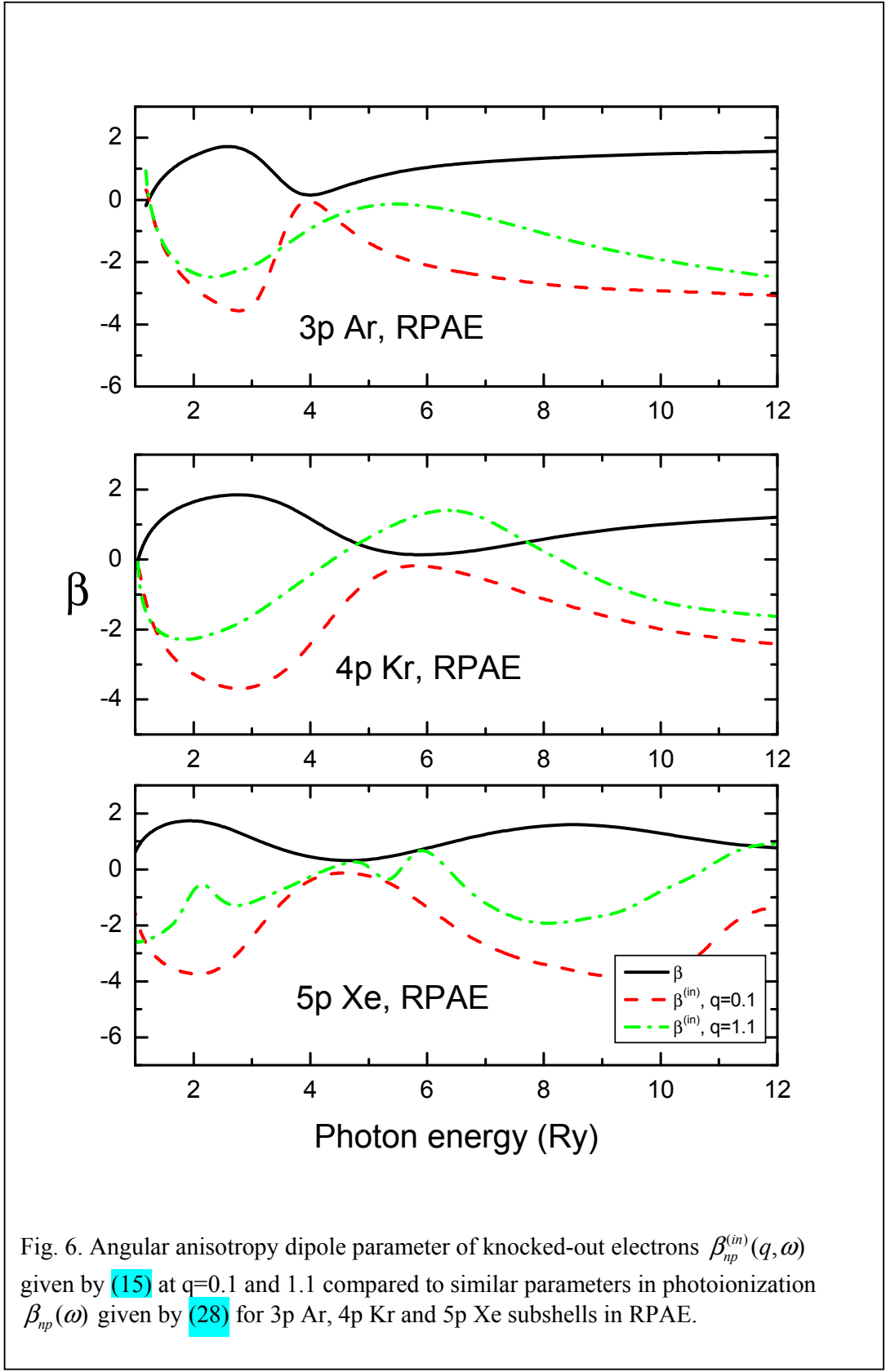

Fig. 6. Angular anisotropy dipole parameter of knocked-out electrons $\beta_{np}^{(in)}(q,\omega)$ given by (15) at q=0.1 and 1.1 compared to similar parameters in photoionization $\beta_{np}(\omega)$ given by (28) for 3p Ar, 4p Kr and 5p Xe subshells in RPAE.

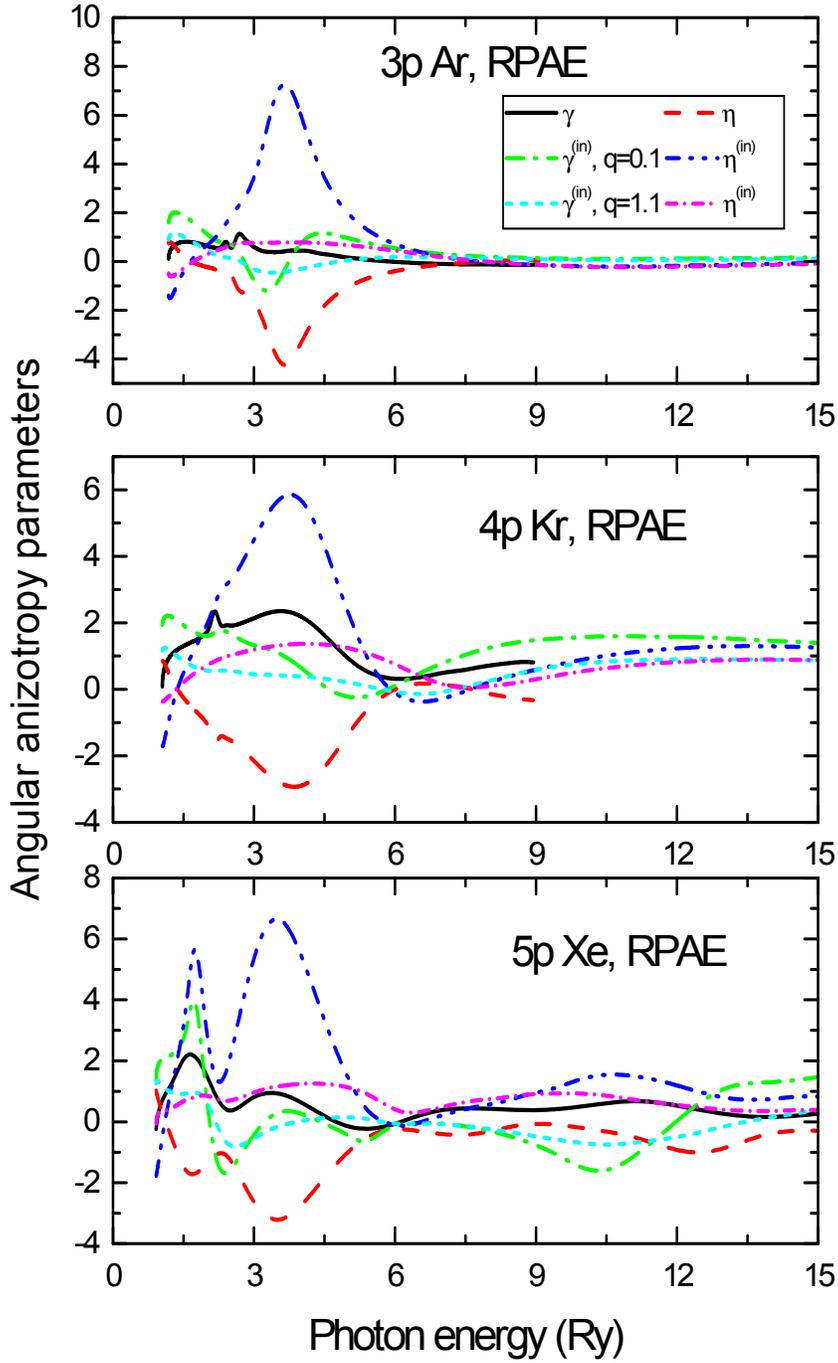

Fig.7. Angular anisotropy parameters of knocked-out electrons $\gamma_{np}^{(in)}(q,\omega)$ and $\eta_{np}^{(in)}(q,\omega)$ given by (15) at q=0.1 and 1.1, compared to similar parameters in photoionization $\gamma_{np}(\omega)$ and $\eta_{np}(\omega)$ given by (29, 30) for 3p Ar, 4p Kr and 5p Xe subshells in RPAE.

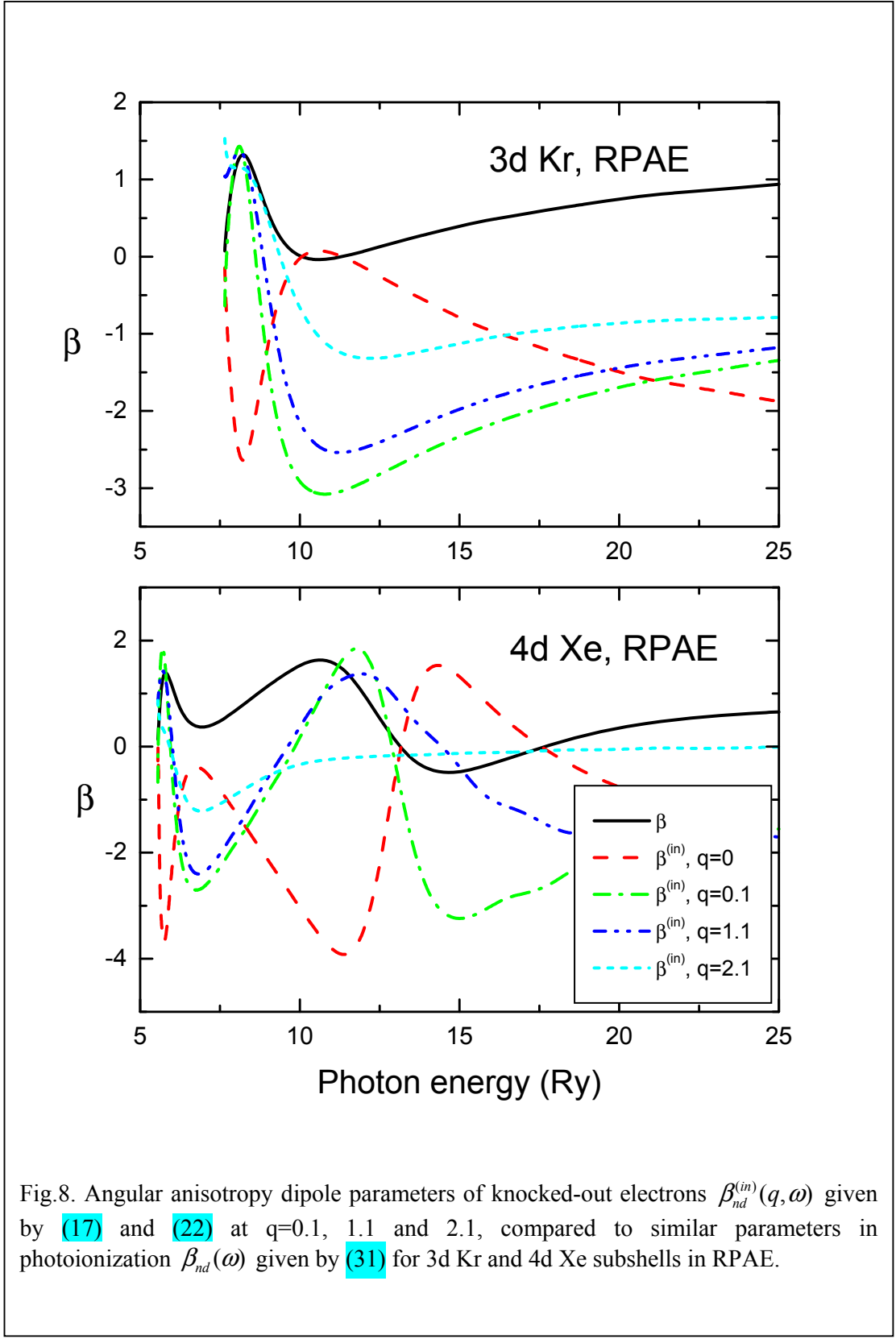

Fig.8. Angular anisotropy dipole parameters of knocked-out electrons $\beta_{nd}^{(in)}(q,\omega)$ given by (17) and (22) at q=0.1, 1.1 and 2.1, compared to similar parameters in photoionization $\beta_{nd}(\omega)$ given by (31) for 3d Kr and 4d Xe subshells in RPAE.

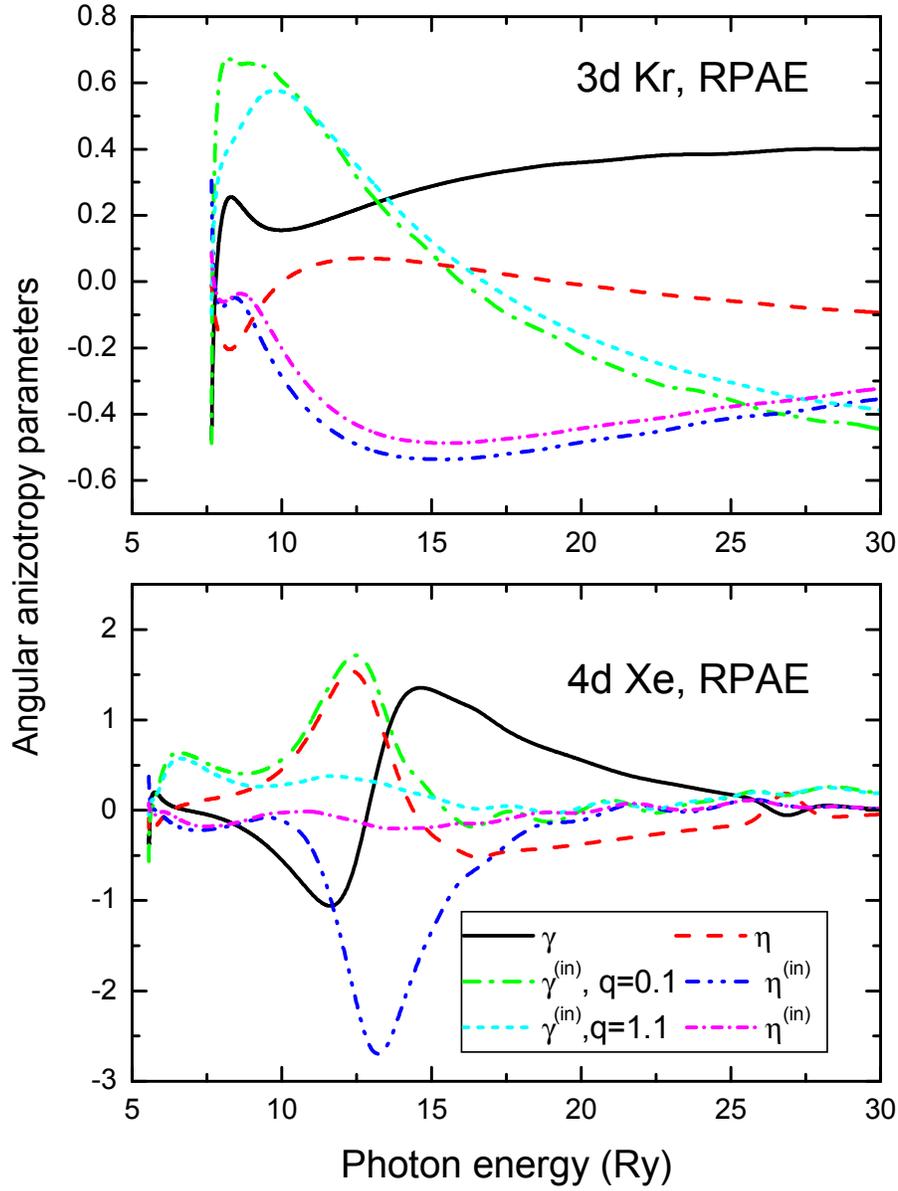

Fig.9. Angular anisotropy non-dipole parameters of knocked-out electrons $\gamma_{nd}^{(in)}(q,\omega)$ and $\eta_{nd}^{(in)}(q,\omega)$ given by (17) at q=0.1 and 1.1, compared to similar parameters in photoionization $\gamma_{np}(\omega)$ and $\eta_{np}(\omega)$, given by (29, 30) for 3d Kr and 4d Xe subshells in RPAE.

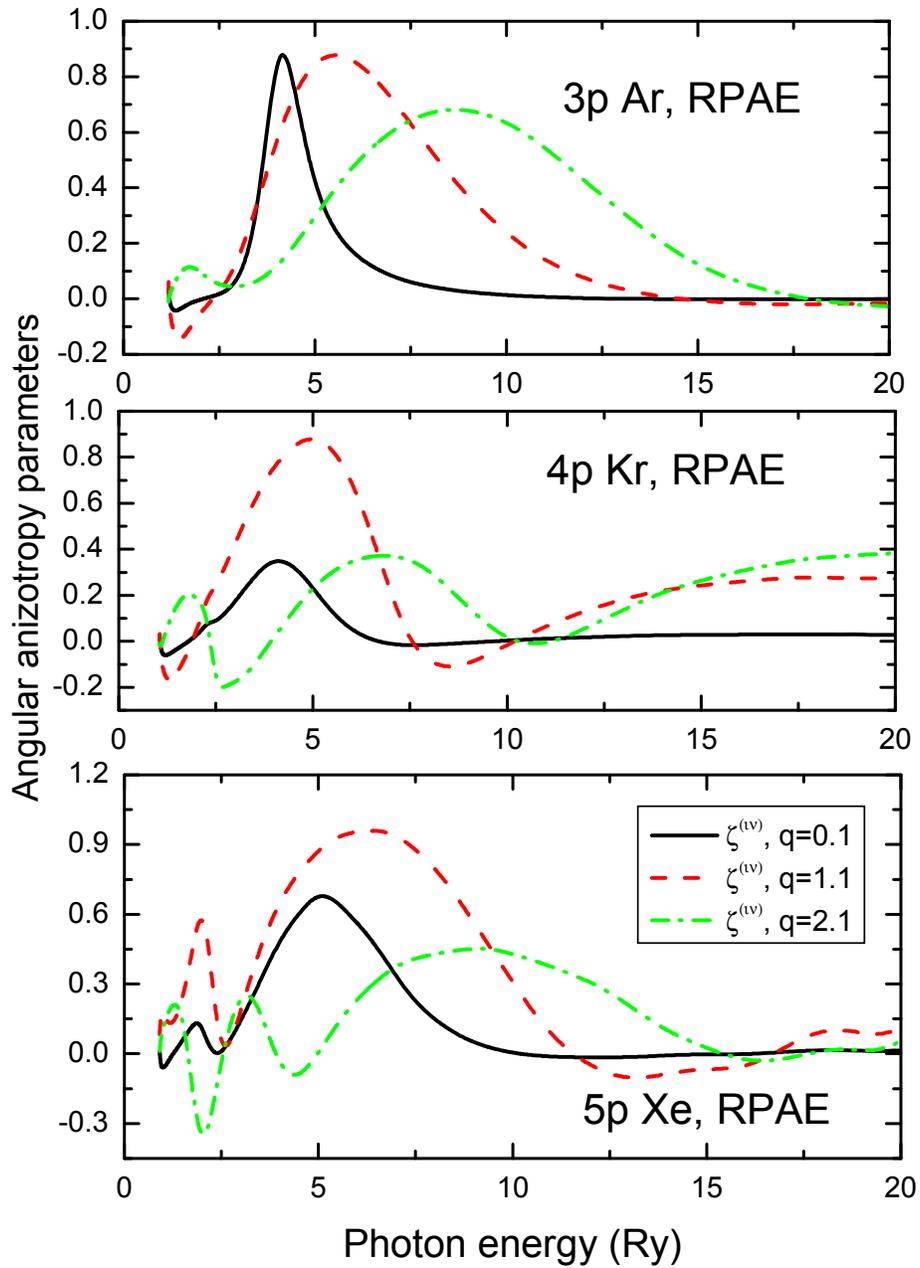

Fig.10. Angular anisotropy non-dipole parameter – coefficient of the fourth Legendre polynomial in the angular distribution of the knocked-out electrons $\zeta_{nl}^{(in)}(q,\omega)$, calculated using (15). The results are presented for 3p Ar, 4p Kr, 5p Xe subshells at q=0.1, 1.1 and 2.1.

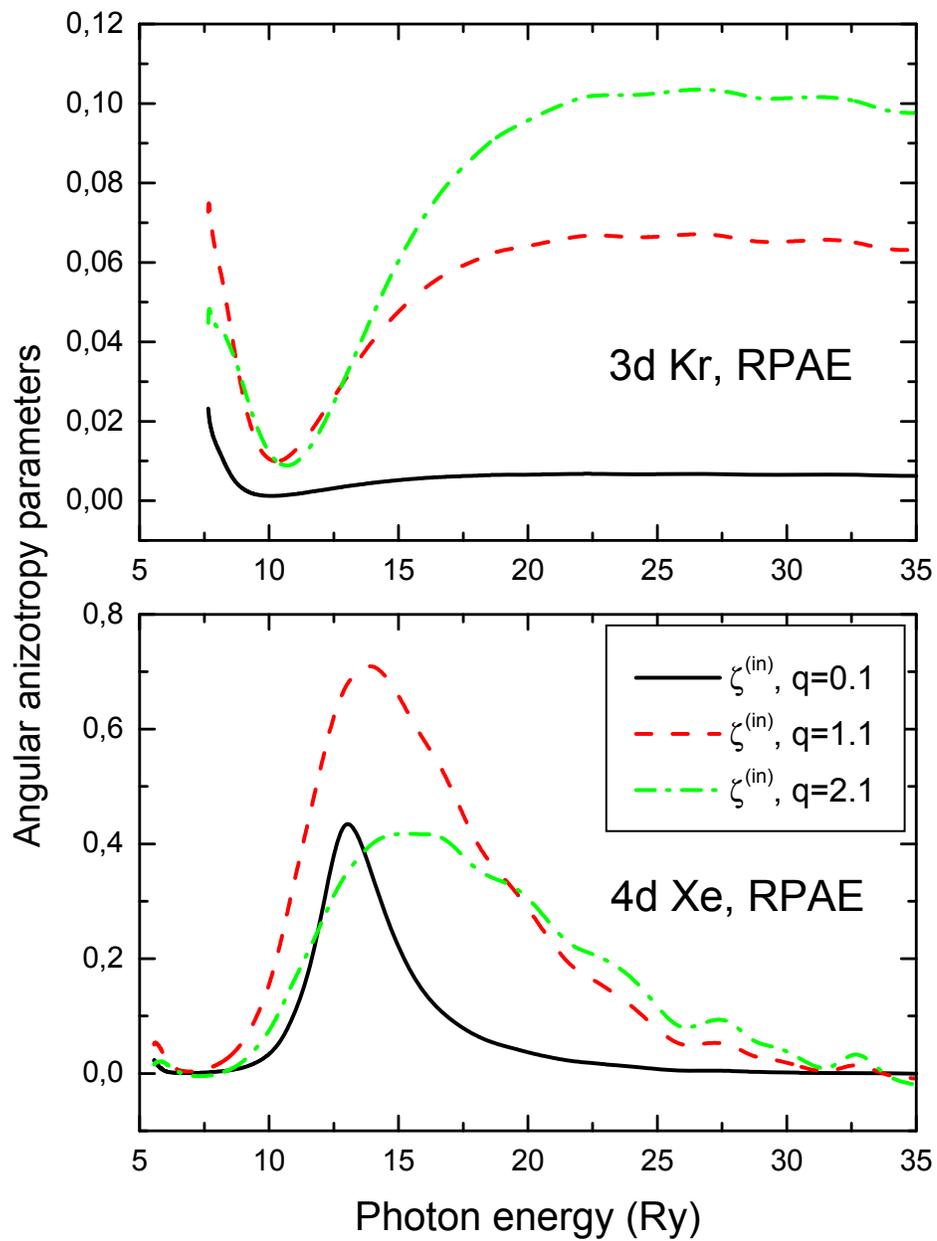

Fig.11. Angular anisotropy non-dipole parameter $\zeta_{nl}^{(in)}(q,\omega)$ – coefficient of the fourth Legendre polynomial in the angular distribution of the knocked-out electrons, calculated using (17). The results are presented for 3d Kr and 4d Xe subshells at q=0.1, 1.1 and q=2.1.